\documentclass[english,aps,pra,reprint,noshowpacs,superscriptaddress]{revtex4-1}   
\usepackage[T1]{fontenc}	
\usepackage[latin9]{inputenc}	
\usepackage{geometry}		
\usepackage{graphicx}
\usepackage[above,below]{placeins}	
\usepackage{times} 
\usepackage{graphicx}
\usepackage{dcolumn}
\usepackage{bm}
\usepackage{tikz}
\usetikzlibrary{shapes.geometric, arrows}

\begin{document}
\title{Surprise! Shifting students away from model-verifying frames in physics labs}
\author{Emily M. Smith}
\author{Martin M. Stein}
\author{N.G. Holmes}
\affiliation{Laboratory of Atomic and Solid State Physics, Cornell University, Ithaca NY 14853}
\begin{abstract}
Framing affects how students interpret, approach, and accomplish tasks. Little is known, however, about how students frame tasks in physics labs. During the first lab of a sequence designed to teach students about modeling and critical thinking with data, students test a simple model of a pendulum that breaks down with improved measurements. Using in-lab video and follow-up interviews, we identified students' frequent use of a model-verifying frame that substantially interferes with the instructional goals. We present a case study analysis of two students who approach the lab with a model-verifying frame, engage in problematic behaviors including questionable research practices, but later shift their frames to accommodate goals aligned with instructional intention. As instructors transition their instructional labs to open-inquiry experiences, an activity that directly challenges the model-verifying frame may be productive for shifting students away from this problematic frame to one that supports their engagement in authentic experimentation.

\end{abstract}

\maketitle

\vspace{-0.5cm}
\section{Introduction}

\vspace{-0.25cm}
Students enter their physics courses with expectations about their role in the classroom. These expectations affect students' decisions and how they choose to approach and engage in activities. Students' expectations and resulting orientations in classroom activities constitute their \emph{frames}~\cite{Framing}. In this paper, we describe a problematic \emph{model-verifying} frame and how challenging this frame may shift students to more productive frames when labs are designed to teach experimentation processes. 

Many labs emphasize physics content; students are expected to verify that physics holds ``true'' in the lab. However, research indicates that there is no measurable added value to emphasizing physics content in the lab~\cite{Holmes15,Holmes17}. Instead, studies point to using the lab as a productive space to teach students other aspects of physics such as experimentation skills and nature of measurement~\cite{Labs1,Labs2,Labs3}. As lab curricula transition to teaching students experimentation skills, students may need to frame labs in new, and perhaps unexpected, ways. Unfortunately, previous lab experiences may influence how students initially frame their physics labs---often as activities to verify models~\cite{Lippman03}. This framing of labs possibly contributes to students' judgment of data as valid when it agrees with theory~\cite{Hu18}. If students frame the physics lab as a verification task, then this may inhibit their ability to see beyond physics content and hinder their learning about the nature of scientific experimentation.

Identifying instructional strategies to shift students' framing of labs may lower the barrier to learning about experimentation such as model evaluation. Here, we report on two students' experiences with an activity that challenged their model-verifying frame and, later, allowed them to reflect on the purpose of a physics lab. Our preliminary case study details two students' shared model-verifying frame as they worked together during lab, the various ways this frame was problematic in the activity, and how, in their perspectives, that experience shifted their ideas about the purpose of and how to engage in labs. This preliminary analysis is part of a larger study that aims to evaluate how students transition to labs that focus on model evaluation.

\vspace{-0.5cm}
\section{Instructional context}

\vspace{-0.25cm}
The labs were part of an honors, calculus-based, introductory mechanics course that contains special relativity and incorporates vector calculus. The activity described in this paper was part of the first lab of the semester. Students completed the lab across two weeks, and, here, we focus on students' experiences during the second week. The lab is intended to teach students about iterating to improve measurements and using statistical tools to evaluate data and inform experimentation decisions. During the first week, through a series of invention tasks~\cite{Schwartz10}, students developed a statistic to compare the distinguishability of two measurements within units of uncertainty, referred to as $t^\prime$~\cite{Labs3}. They were provided with guidelines:

\vspace{-0.3cm}
\begin{center}
$t^\prime<1$, Indistinguishable measurements\\
$1<t^\prime<3$, Inconclusive\\
$t^\prime>3$, Distinguishable measurements
\end{center}

\vspace{-0.25cm}
During the second week, students began comparing the period of a pendulum at $10^\circ$ and $20^\circ$  and iterated to improve measurements. With improved measurements (e.g., additional trials, increasing oscillations per trial), $t^\prime$ increases, and it is possible to experimentally distinguish the periods at $10^\circ$ and $20^\circ$ with $t^\prime>3$---i.e., the small-angle approximation in the model breaks down. While many groups were able to extend their investigation to test other effects (e.g., angular dependence, length of string), the pair in this paper did not extend their investigation.

The lab handout provided a model for the period of the pendulum: $T=2\pi\sqrt{{L}/{g}}$. The language (e.g., test, evaluate whether) around this model was not intended to imply that this is necessarily an appropriate model for the situation. The handout prompted students to iterate to improve their measurements (i.e., reduce uncertainty). Instructors emphasized that the labs were designed to engage students in experimentation processes rather than reinforcing physics concepts.

\vspace{-0.25cm}
\section{Methods}

\vspace{-0.25cm}
This study is a preliminary case study investigation of students' model-verifying frames. The data uses classroom video of two students, Ali and Ben, during the second week of the lab activity and audio recordings of individual interviews six months later about their experience completing the activity. Ali and Ben were first semester college students and were highly prepared for introductory mechanics having taken college-level physics in high school.

We video-recorded six groups' behaviors and conversations within two lab sections. To identify students' framing of lab activities, our preliminary analysis of the video began by identifying, in all groups, instances when students articulated their expectations about the results and purpose of the experiment. We selected Ali and Ben for the preliminary study because they were explicit about their expectations at the outset. Three of the five other groups implied that they initially understood the purpose to be verifying the model but only Ali and Ben began with a discussion of their shared understanding of the task as verification of the model, a model-verifying frame. 
After broadly categorizing students' expectations in all video-recorded groups, one of us (E.M.S.) returned to identify specific instances where Ali and Ben's expectations were explicitly and implicitly conveyed through their conversations, decisions, and behaviors. Then, one of us (N.G.H.) watched the recording to search for contradictory evidence to the examples and interpretations provided in this paper.

To investigate what Ali and Ben believed they learned from the lab, we conducted individual semi-structured interviews six months after the activity. During the interview, they received their group's lab notes and were asked to describe why their group chose to make the recorded decisions. After talking through their notes, each student was asked to reflect on what they learned from the activity. We designed the prompts to target students' reflections on their group's decision-making processes in retrospect and whether they believe that they shifted their framing of lab activities in subsequent labs and/or courses. We did not expect students to accurately recall their in-lab decisions and discussions. 

\vspace{-0.25cm}
\section{Model-verifying frame\label{sec:framing}}

\vspace{-0.25cm}
In this section, we have time ordered a selection of events to demonstrate how Ali and Ben negotiate frames and how their shared model-verifying frame influences their behaviors throughout the two hour lab. 

\vspace{-0.5cm}
\subsection{Definition}

\vspace{-0.25cm}
In this paper, we claim that students exhibit a model-verifying frame when they understand the purpose of a lab to be verifying or demonstrating that a model provided to them (through lecture, lab materials, etc.) holds ``true'' in the lab. Due to the preliminary status of this work, our definition of a model-verifying frame is broad and described at a large grain-size but may manifest in distinct ways among individuals and groups who understand labs as verifying activities. Several frames interact with Ali and Ben's model-verifying frame but their model-verifying frame is dominant and persists throughout the entire lab period. While this frame may be productive for many labs that aim to reinforce physics content, it is problematic in labs that incorporate aspects of model evaluation. 

Ali and Ben explicitly frame the activity as verifying the provided model from the outset of the activity. As they are setting up the pendulum prior to the lab session starting, they discuss that this week they will be comparing different angles and that ``\emph{the purpose of the lab is to prove that the period is not affected by the angle.}'' They check the lab handout and do not find a procedure, which leads them to discuss that ``\emph{going about it with the purpose of proving that they're not affected is a good idea.}'' Immediate mutual acceptance of the purpose of the lab suggests they share a common model-verifying frame. 

After their first iteration, Ali and Ben set out to improve their measurements---encouraged by lab instructions to decrease the uncertainty of their data---and their results become puzzling; additional trials increase $t^\prime$. They begin to wonder ``\emph{why is [$t^\prime$] so high?}'' After checking their calculations, they continue to reflect ``\emph{so why's it so big? [$t^\prime$] is supposed to be less than one. That's so odd.}'' The results, which do not conclusively verify the model, are perplexing to the students due to their model-verifying frame.  

\vspace{-0.5cm}
\subsection{Problematic persistence of the frame}

\vspace{-0.25cm}
Despite these puzzling initial results, Ali and Ben persist in a model-verifying frame for the remainder of the lab. However, their model-verifying frame prevents Ali and Ben from participating in the activity as intended. We provide examples that highlight how the frame is problematic for the students' productive involvement in the lab activity.

\vspace{-0.5cm}
\subsubsection{Ignoring instructional hints \& reluctance to search for contradictory evidence}

\vspace{-0.25cm}
Ali and Ben's model-verifying frame prevents them from understanding the conversations they have with the lab instructor, which is a source of frustration. Their frustration appears to emerge due to discussions with the instructor that conflict with their model-verifying frame. The instructor consistently encourages them to seek out other groups' results that contradict their own and hints at questioning the theoretical model, however, the students are reluctant to act on this advice. These episodes illuminate the conflict of instructional messaging with the model-verifying frame.

As Ali and Ben iterate to find that $t^\prime$, again, increased and is in the inconclusive range, the instructor checks on their progress. They explain to the instructor that ``\emph{the [$t^\prime$] is worse.}'' The instructor attempts to understand the situation and their reasoning by probing ``\emph{so your [$t^\prime$] is growing as you acquire more data? So what do you think it should do?}'' Ali and Ben communicate to him that the periods should not be distinguishable, conveying their model-verifying frame.

To push the students to think beyond verifying the model, the instructor points to the model and asks ``\emph{so this is a theoretical equation, right? Are there any things that might make them distinguishable?}'' Ali and Ben respond with several concerns about data collection rather than critically examining the model, as the instructor hinted. This response suggests that because Ali and Ben persist in the model-verifying frame they ignore instructional hints to critique the theoretical model. The model holds greater authority than the instructor and, possibly, experimental methods and results.

In contrast to interactions with the instructor, Ali and Ben's interactions with peers validate their model-verifying frame, which reinforces their reluctance to search for contradictory evidence. Ben engages with Cara, from another group, who shares a model-verifying frame. Cara asks ``\emph{how is your $t^\prime$ value?}'' to which Ben replies that the statistic is ``\emph{kind of a pain in the ass. Ours was great but then the [instructor] asked `why don't you try something to make it even better?' So we start making it better and it just keeps going [higher].}''  Cara comments that Ali and Ben's $t^\prime$ is not too far away from statistically indistinguishable periods, reinforcing a model-verifying frame. Cara explains that when measuring from the top of the swing her group found the periods were distinguishable. However, she explains that the result ``\emph{is probably due to like human error than due to like actual scientific merit.}'' Cara creates an unsupported explanation for discrepancies between the model and experimental results rather than seeking additional data.

After Cara leaves, Ali appears willing to reevaluate their objectives and asks Ben ``\emph{what are we supposed to like find in the end?}'' Ali considers whether the goal of the lab is to verify the model, possibly opening up to a search for contradictory evidence. However, Ben replies that ``\emph{this whole lab is just to see how we do labs; I don't think it's actually [a lab].}'' Ben confirms that he believes the unstated goal of the lab is to demonstrate how they should behave in future labs; they should aim to confirm models. 

Due to their insistence that the task is about verifying the model, Ali and Ben are unable to act on the support they receive and disregard guidance to seek out contradictory evidence. The instructor returns to discuss plans and the students convey their concern that additional trials will ``\emph{make $t^\prime$ worse,}'' recognizing that improved precision increases $t^\prime$. Pushing them to seek out contradictory evidence, the instructor suggests that the students try another angle or talk to another group with different results. However, Ben directs his question to a student who also has obtained inconclusive values of $t^\prime$ and comments that ``\emph{it's not good},'' confirming Ali and Ben's expectations about results. The instructor attempts to support their learning about the experimentation process by providing possible routes for investigating the discrepancy between the model and results. However, the students return to discussing their experiment, again brushing off instructional hints to seek contradictory evidence.   

\vspace{-0.5cm}
\subsubsection{Engaging in questionable research practices}

\vspace{-0.25cm}
As the lab progresses, Ali and Ben's model-verifying frame leads them to engage in questionable research practices beyond their reluctance to seek out contradictory evidence. As they plan their final iteration, they critically evaluate the formula for calculating the $t^\prime$ statistic. In doing so, they evaluate the features of the experiment that decrease the magnitude of the statistic; they decide to manipulate the experiment to obtain their desired result.

To decrease $t^\prime$, Ali and Ben lengthen the string of the pendulum: They expect the periods to be longer and intend to test whether lengthening the string ``\emph{will help or hurt [their] data.}'' They also choose to revert to measuring the period of single swings because they acknowledge that their initial procedure provided results that demonstrated that the periods are the same. Their earlier decisions to improve their measurements were motivated by decreasing the standard deviation of data sets. However, their desire to verify the model prevails; they view improved measurements as those that verify the model. In doing so, they purposefully decrease the precision of their measurements to obtain a lower, and more desirable, $t^\prime$ value---a questionable practice in research. Interestingly, they record their decisions in their lab notes indicating that they did not believe these decisions were questionable and, perhaps, believed their instructor would evaluate their success in verifying the model. 

\vspace{-0.5cm}
\subsection{Challenging the model-verifying frame}

\vspace{-0.25cm}
Through a whole class discussion at the end of the lab, Ali and Ben's model-verifying frame is directly challenged when other groups explain that they uncovered the small angle approximation in the model through improved measurements of the periods at $10^\circ$ and $20^\circ$. When a student discusses the small angle approximation in the model, Ali appears concerned and both students recheck the model of the period. Ali and Ben's concern appears to result from learning that the model broke down, a surprising and unexpected event: The purpose of the lab was not to verify the model. But the students do not act on this information and submit their notes without additions or changes.

\vspace{-0.25cm}
\section{Shifting frames in lab activities\label{sec:interviews}}

\vspace{-0.25cm}
In interviews six months later, Ali and Ben reflected on their in-lab decisions as they read through their group's lab notes and responded to follow-up questions reflecting on what they learned during the activity.

\vspace{-0.5cm}
\subsection{Ali's shifted framing of labs}

\vspace{-0.25cm}
Ali's responses during the interview suggest that she continues to frame labs, in part, as model-verifying activities but with the added possibility that a model may have embedded assumptions. As Ali read through her group's lab notes she paused to reflect that their decision to revert back to single swings was interesting ``\emph{'cause that's kind of like going backwards in our method.}'' However, she claims that their decisions were ``\emph{happening because [they] forgot about small angle approximation.}'' Ali realizes they backtracked on improving measurements during the lab but explains that this happened because they forgot an assumption in the model. 

Ali also describes that their sustained model-verifying frame restricted the possible conclusions that they could draw from their data. She reflects that with the most precise dataset: ``\emph{I feel like we would've been able to like prove like `oh yes, like it is different at different angles' but we were trying to prove the opposite and then our data was (sic) like not supporting that.}'' Ali considers possible alternative decisions such as recognizing that ``\emph{oh our data's very precise but it's not like accurate with what we're like talking about so maybe we should refocus what we should be looking for.}'' This suggests that Ali recognizes a different frame would have been productive for fully engaging in the task. 

From this experience, she learned that ``\emph{it's important to perhaps like when data is like not leading in the direction that you were thinking of to like explore why it is doing that instead of... doing the same exact thing and hope that it goes... in your favor,}'' suggesting that she learned to incorporate critical thinking about disagreements between data and models rather than merely verifying models. Ali's discussion of exploring unexpected results suggests she has extended her framing of labs though the primary purpose continues to be verifying models, perhaps with unexpected assumptions. 

\vspace{-0.5cm}
\subsection{Ben's shifted framing of labs}

\vspace{-0.25cm}
Throughout the interview, Ben's statements suggest he believes that he shifted his framing of lab activities due to the experience of comparing pendulum periods. Ben reflects that he learned to ``\emph{look at the data on its own and look past any preconceived ideas about what the outcome should be.}'' He recognizes that he expected the periods to be the same at all angular displacements and the experience prompted him to think beyond his expectations. Over the semester, the labs provided him with understanding of how to ``\emph{base [his] conclusions totally off the data and how reliable the data was (sic) and what terminology [he] could use in [his] conclusions and how it, how [he could] quantitatively express how, how correct or how well the data supports the hypothesis.}'' 

Ben's new frame(s) in lab activities accommodate inconclusive and contradictory data as acceptable within an experiment. Unprompted, Ben reflects on what he would now conclude about his group's data from comparing pendulum periods. He ``\emph{would be more inclined to say... `you know what, I think this data is inconclusive. I don't know if we can necessarily say that this is true',}'' but offered that at the time that he was thinking ``\emph{obviously we're given this equation and it doesn't depend on $\theta$, so... that data just has some source of error in it, and we're fine.}'' His ability to explain how he would choose to make conclusions from his data now that he has learned these lessons suggests that he has significantly changed his perception of acceptable behavior in lab in ways that align with instructional intention. 

\vspace{-0.25cm}
\section{Discussion}

\vspace{-0.25cm}
Ali and Ben initially frame the lab as a verification task and are unable to engage in model evaluation as intended. Their model-verifying frame causes them to ignore contradictory evidence, brush off instructional hints, and manipulate the experiment to obtain desired results. Their behaviors suggest that their previous experiences with labs---presumably with objectives centered around verifying models---influenced their engagement in questionable research practices. We have identified evidence of questionable research practices in students' lab notes at three institutions~\cite{Stein18} and, in future work, plan to investigate whether students' prior experiences in verification labs motivate questionable research practices early in students' transition to experimentation-focused labs.

From the case of Ali and Ben, we expect there are several types of model-verifying frames marked by beliefs about models, statistics, instructors, instructions, and experimental results in labs. Ali and Ben appear to share beliefs about instructional labs including that models provide truth and that statistics confirm truth. Through similar analyses of other groups that exhibit a model-verifying frame, we intend to identify beliefs associated with this frame and the ways that this activity challenges these beliefs. In doing so, we aim to understand whether students' beliefs, and framing of labs, evolve in subsequent lab activities. 

Instructional labs may provide an authentic opportunity to teach students about ethical and rigorous research by intentionally shifting them away from a model-verifying frame. Lab curricula that aim to have students engage in authentic practices need to encourage students to shift frames, otherwise students may fall into unproductive, and possibly unethical, behaviors. Ali and Ben opened to the possibility that labs are not merely verification tasks, and attributed their shifted perspectives to their experiences with this activity. This suggests that developing activities that require students to confront their model-verifying frame may rapidly shift students away from this problematic frame to one that supports their engagement in authentic experimentation. 

\vspace{-0.5cm}
\acknowledgments{\vspace{-0.25cm}This work was partially supported by Cornell University's College of Arts \& Sciences Active Learning Initiative.}

\end{document}